\documentstyle[aps,multicol,epsf]{revtex}
\begin{document}

\title{Predictive Information}
\author{William Bialek$^1$ and Naftali Tishby$^{1,2}$}
\address{$^1$NEC Research Institute, 4 Independence Way, Princeton, New
Jersey 08540\\
$^2$Institute for Computer Science \& Center for Neural
Computation, Hebrew University, Jerusalem, Israel}

\date{\today}
\maketitle

\begin{abstract}
Observations on the past provide some hints about what will
happen in the future, and this can be quantified using information
theory.  The ``predictive information'' defined in this way
has connections to measures of complexity that have been proposed
both in the study of dynamical systems and in mathematical
statistics.  In particular, the predictive information
diverges when the observed data stream allows us to learn an increasingly
precise model for the dynamics that generate the data, and the structure of
this divergence measures the complexity of the model.  We argue that
divergent contributions to the predictive
information provide the only measure of
complexity or richness that is consistent with certain plausible
requirements.
\end{abstract}


\begin{multicols}{2}

%
%
%
%
%
%
%
%

There is obvious interest in having practical algorithms
for predicting the future, and there is a correspondingly
large literature on the problem of time series extrapolation \cite{extrap}. But prediction is more (and less) than 
extrapolation---we might be able to  predict,  for example, the chance of rain
in the coming week even if we cannot extrapolate the trajectory of temperature
fluctuations. In the spirit of its thermodynamic origins, 
information theory \cite{shannon} characterizes the potentialities and limitations of all
possible prediction algorithms, as well as unifying the analysis of
extrapolation with the more general notion of predictability.  Specifically, we
define a quantity---the {\em predictive information}---that measures how
much our observations of the past can tell us about the future.  The predictive
information characterizes the world we are observing, and we shall see that
this characterization is close to our intuition about the complexity of the
underlying dynamics.

Imagine that we
observe a stream of data over the period from $t=-T$ to $t=0$; this constitutes
our past.  For simplicity, let the future extend forward also for a
time
$T$.  Let us call the data we observe during the past $X_{\rm past}$, and the
data we will observe in the future $X_{\rm future}$.  The usual problem 
is to guess the values of $X_{\rm future}$ from knowledge of
$X_{\rm past}$, but this is too specific; only certain
features of the data stream are predictable, and even these features may 
be predictable only in a statistical sense. Different kinds of prediction are often
treated as  different problems, and when we assess the quality of these predictions we seem
forced to use different metrics in the different cases.  
Information theory allows us to treat the different notions of prediction on
the same footing. 

Even before we look at the data,
we already know that certain futures are more likely
then others, and we can summarize this knowledge by a `prior' probability
distribution for the future,
$P(X_{\rm future})$. 
Our observations on the past lead us to a new, more tightly concentrated
distribution, the distribution of futures conditional on the past data,
$P(X_{\rm future}| X_{\rm past})$. Different kinds of predictions can be seen as
different slices through or averages over this conditional distribution. The greater concentration
of the conditional distribution can be quantified directly by the fact that it has a smaller
entropy than the prior distribution, and this reduction in entropy is Shannon's
definition of the information that the past provides about the future
\cite{shannon}.  We can write the average of this  predictive information
as
\begin{eqnarray}
I_{\rm pred} (T) 
&=& 
\Bigg\langle \log_2 \left[
{{P(X_{\rm future} , X_{\rm past})}
\over{P(X_{\rm future})P( X_{\rm past})}}\right]\Bigg\rangle \\
&=&
-\langle\log_2 P(X_{\rm future})\rangle
-\langle\log_2 P( X_{\rm past})\rangle\nonumber\\
&&\,\,\,\,\,\,\,\,\,\,
-\left[-\langle\log_2 P(X_{\rm future} , X_{\rm past})\rangle\right] .
\label{ents}
\end{eqnarray}
Each of the terms in Eq. (\ref{ents}) is an entropy. 
If we have invariance under time translations, then
the entropy of the past data depends only
on the duration of our observations, so we can
write
$ -\langle\log_2 P( X_{\rm past})\rangle = S(T) $,
and by the same argument
$-\langle\log_2 P( X_{\rm future})\rangle = S(T) $.
Finally, the entropy of the past and future taken together is the entropy of
observations on a window of duration $2T$, so that
$ -\langle\log_2 P(X_{\rm future} , X_{\rm past})\rangle
= S(2T)$.
Putting these equations together we obtain the basic relation between 
predictability and
the time dependence of the entropy,
\begin{equation}
I_{\rm pred}(T) = 2S(T) - S(2T) .
\label{IpredandST}
\end{equation}

The entropy  is an extensive quantity,
so that
$\lim_{T\rightarrow\infty} {{S(T)}/ T} = {\cal S}.$
In the same way that the entropy of a
gas at fixed density is proportional to the
volume, the entropy of a time series is
(asymptotically) proportional to its duration.
This entropy is also the minimum number of bits required to give a complete
description of the past data. 
But  from Eq. (\ref{IpredandST}) any extensive component of the
entropy cancels in the computation of the predictive information: predictability is
associated with  deviation of the entropy from extensivity.  The cancellation of
extensive components means that the
predictive information must be subextensive, 
$\lim_{T\rightarrow\infty} {{I_{\rm pred}(T)}/ T} = 0.$
As a result, of the total information we have taken in by observing
$X_{\rm past}$, only a vanishing fraction 
is of relevance to predicting the future:
\begin{equation}
\lim_{T\rightarrow\infty} 
{{\rm Predictive\ Information}
\over{\rm Total\ Information}}
=
{{I_{\rm pred} (T)}
\over {S(T)}}
\rightarrow 0.
\label{chuck}
\end{equation}
In this precise sense, most of what we observe is irrelevant to the problem of
predicting the future.

Qualitatively, we expect the predictive information to behave in one of
three ways for large values of the time $T$.  One possibility is that, no
matter how long we observe, we learn only a finite amount of information about
the future, so that $\lim_{T\rightarrow\infty} I_{\rm pred} (T) =$ constant.
This situation prevails when even the best possible predictions are controlled
only by the immediate past, so that the correlation times of the
observable data are finite.  Alternatively, the predictive information can be small
because the dynamics are too regular:  for a purely periodic system, complete prediction is
possible once we know the phase, and if we sample the data at discrete times this a
finite amount of information; longer period orbits are intuitively more complex and also
have larger $I_{\rm pred}$.  In physical
systems we know that there are critical points where correlation times become infinite, so that
optimal predictions will be influenced by events in the arbitrarily distant past. Under these
conditions  the predictive information can grow without bound as $T$ becomes large; for
many systems the divergence is logarithmic, 
$I_{\rm pred} (T\rightarrow\infty) \sim\mu \ln T$.
Finally it
is possible that $I_{\rm pred} (T\rightarrow\infty)
\propto T^\alpha$.

Imagine that we observe $x(t)$ at a series of discrete times $\{t_{\rm n}\}$, 
and that each time point we find the value $x_{\rm n}$.
Then we can always write
the joint  distribution of the
$N$ data points as a product, 
\begin{equation}
P(x_1 , x_2 , \cdots , x_N )
= P(x_1 ) P(x_2 | x_1) P(x_3 | x_2 , x_1) \cdots .
\label{joint}
\end{equation}
For  Markov processes,   what we observe at $t_{\rm n}$ depends only on events at
the previous time step $t_{\rm n-1}$, so that
\begin{equation}
P(x_{\rm n} | \{x_{\rm 1\leq i \leq n-1}\}) = P(x_{\rm n} | x_{\rm n-1}) ,
\end{equation}
and hence the predictive information reduces to
\begin{equation}
I_{\rm pred} = \Bigg\langle \ln\left[
{{P(x_{\rm n}|x_{\rm n-1})}\over{P(x_{\rm n})}}
\right] \Bigg\rangle .
\end{equation}
The maximum possible predictive information is 
the entropy of the distribution of states at one time step,
which in turn is bounded by the logarithm of the number of accessible states.
To approach this bound the system must maintain memory for a long time,
since the predictive information is reduced by the entropy of the
transition probabilities.
Thus systems with more states and longer memories have 
larger values of $I_{\rm pred}$.

Consider next a time
series of pairs $(x_1 , y_1)$, 
$(x_2 , y_2)$, $\cdots$,  $(x_N , y_N )$.  The points $x_{\rm n}$ are chosen independently and at
random from some $P(x)$,   while the $y_{\rm n}$ are noisy examples of
a function
$f(x)$,  $y_{\rm n} = f(x_{\rm n }) + \eta_{\rm n}$, with the $\eta_{\rm n}$  
chosen, for instance, from a Gaussian distribution.  Let us assume that the function $f(x)$ can be
written as sum of 
$K$ basis functions $\phi_1(x) , \phi_2 (x), \cdots ,\phi_K(x)$ with unknown
coefficients $\alpha_\mu$.  Then the joint distribution of  $\{ x_{\rm n} , y_{\rm n}\}$
is given by
\begin{eqnarray}
P(\{ x_{\rm n} , y_{\rm n}\} )
&=& \left[ \prod_{\rm n =1}^N P(x_{\rm n})\right]
{1\over{(2\pi\langle\eta^2\rangle)^{N/2}}}
\nonumber\\
&&\,\,\,\,\,\times
\int d^K \alpha
P(\{\alpha_\mu\}) 
\exp ( - \chi^2/2 ),\\
\chi^2 &= &{1\over{\langle \eta^2\rangle}}
\sum_{\rm n=1}^N \left|
y_{\rm n} - \sum_{\mu =1}^K \alpha_\mu \phi_\mu(x_{\rm n})\right|^2  .
\end{eqnarray}
In the limit $N \rightarrow \infty$  the integral over the parameters $\alpha_\mu$
can be done in a saddle point approximation \cite{vijay,ilya}, and in this approximation the
entropy of the distribution $P(\{x_{\rm n} , y_{\rm n}\})$ has an extensive term proportional to
$N$ but also a leading subextensive term $\sim (K/2)\ln N$.
The result is the same for a broad
class of time series in which the data are described by $K$ parameter models with unknown
parameters \cite{ilya}: the predictive information is $I_{\rm pred} = (K/2)\ln N$ and is equal
to the information that the data provide about underlying dynamical model.

Rather than being described by a finite number of parameters, it is possible
that the functional relations embedded in
the data $\{x_{\rm n} , y_{\rm n}\}$ reflect 
an arbitrary smooth function \cite{learnprl}.
As we observe more of the time series we expect to
give a more and more sophisticated description of this underlying function, in
effect allowing the number of parameters in our description to increase with
time.   This suggests that the predictive information, which is proportional to
the number of parameters in the finitely parameterizable case, will grow more
rapidly with
$N$ in the nonparametric setting \cite{ilya}.  Similarly, when we examine written texts on the
scale of tens of letters, we learn about the rules for combining letters into
words,  but if we look at hundreds of letters we also learn about the rules for
combining words into phrases, and so on:
longer texts teach us about an
increasing number of different things,   rather than giving us more precise knowledge about a
fixed number of rules or parameters.  Statistical analyses of long texts suggest that their
entropy has a large subextensive component, and that this component---and hence
the predictive information---is best fit by a power law, so that $I_{\rm pred}
\propto N^{1/2}$ for $N$ letter texts \cite{texts}. This result agrees with a
recent reanalysis \cite{hilberg} of Shannon's classic
experiments on the prediction of English texts by human observers
\cite{ces-english}.

The divergence of the predictive information has an interesting consequence. 
The  average amount of information we
have about the current state of a signal is (asymptotically)
independent of how long we have been watching. 
On the other
hand, if  we live in
a world such that signals have diverging predictive information then the
space required to write down our description grows and grows as we observe the
world for longer peirods of time.  In particular, if we can observe for a very
long time then the amount that we know about the future will exceed, by an
arbitrarily large factor, the amount that we know about the
present \cite{cortex?}.

The examples considered here suggest that the predictive information
corresponds to our intuitive notion of complexity in the incoming data stream:
$I_{\rm pred}$ distinguishes processes that can be described by a finite number
of parameters from those that cannot, and within each
class counts the number of parameters or dimensions that are relevant. The
problem of quantifying complexity is very old \cite{kolmo}. There are two major
motivations.  First, we would like to make precise our impression that some
systems---such as life on earth or a turbulent fluid flow---evolve toward a
state of higher complexity.  Second, in choosing among different models that
describe an experiment, we want to quantify  our preference for simpler
explanations or, equivalently, provide a penalty for complex models that can be
weighed against the more conventional `goodness of fit' criteria.

The construction of complexity penalties for model selection is  a
statistics problem. In this context, Rissanen has emphasized that fitting a
model to data represents an encoding of those data, and that in searching for an
efficient code we need to measure not only the number of bits required to
describe the deviations of the data from the model's predictions (goodness of
fit), but also the number of bits required to specify the parameters of the
model, which he terms the stochastic complexity \cite{rissanen}.  For models
with a finite number of parameters, the stochastic complexity is proportional
to the number of parameters and logarithmically dependent on the number of data
points we have observed,  as found here for the predictive
information.  The connection of stochastic complexity to statistical mechanics
ideas has also been noted by Balasubramanian \cite{vijay}.

The essential difficulty in constructing complexity measures for physical
systems is to distinguish genuine complexity from randomness (entropy).  Several authors have
considered complexity measures related to the mutual information between spatially distant
points, but this is problematic \cite{bennett}.  
Lloyd and Pagels \cite{pagels}
identified complexity (thermodynamic depth)
with the entropy of the state sequences
that lead to the current state, an idea which is clearly in the same spirit as the
measurement of predictive information, but this depth measure does not completely
discard the extensive component of the entropy.  Grassberger has emphasized that the slow approach
of the entropy to its extensive limit is a sign of complexity, and has proposed a function, the
effective measure complexity, that isolates this term in a
form almost equivalent to the predictive information
\cite{grassberger}.  Crutchfield and Young \cite{crutch} have argued that the effective measure
complexity is also related to number of states in the minimal finite state machine that would
simulate the dynamics of the data stream.  For low dimensional dynamical
systems, the effective measure complexity and hence the predictive information
is finite whether the system exhibits periodic or chaotic behavior, but at the
bifurcation point that marks the onset of chaos the predictive information
diverges logarithmically. 
Simulations of specific cellular automaton models that are capable of universal
computation indicate that these systems exhibit a power law divergence of
$I_{\rm pred}$ \cite{grassberger}.

We recall that entropy provides a measure of information that is unique in satisfying certain
plausible constraints \cite{shannon}.  It  
would be attractive if we could prove a similar uniqueness theorem for the 
predictive
information as a measure of the complexity or richness of a time dependent 
signal $x(0 < t <T)$ drawn from a distribution $P[x(t)]$.
As in Shannon's approach, such a measure must obey some constraints:
if there are $N$ equally likely signals, then the measure should be 
monotonic in $N$;
if the signal is decomposable into statistically independent parts then the
measure should be additive with respect to this decomposition; and if the 
signal
can be described as a leaf on a tree of statistically independent decisions 
then
the measure should be a weighted sum of the measures at each branching 
point. For discrete signals
these criteria specify the entropy of the distribution $P[x(t)]$ as a 
unique measure,
but in the case of continuous signals there are ambiguities.  We would like 
to write
the continuum generalization of the entropy,
\begin{equation}
S_{\rm cont} = -\int Dx(t) \,P[x(t)]\log_2 P[x(t)] ,
\end{equation}
but this is not well defined because we are taking the logarithm of a 
dimensionful
quantity.  Shannon gave the solution to this problem:  we use as a measure 
of information
the relative entropy between the distribution $P[x(t)]$ and some reference 
distribution
$Q[x(t)]$,
\begin{equation}
S_{\rm rel} = -\int Dx(t) \,P[x(t)]\log_2 \left({{P[x(t)]}\over{Q[x(t)]}}
\right) ,
\end{equation}
which is  invariant under changes of our coordinate 
system on the
space of signals.  The cost of this invariance is that we have introduced 
an arbitrary distribution
$Q[x(t)]$, and so we really have  a family 
of
measures.  We suggest that, within this family, the appropriate measure of 
complexity is one which obeys further invariance principles.

The reference distribution $Q[x(t)]$ embodies our expectations for the signal $x(t)$;
in particular, $S_{\rm rel}$ measures the extra space needed to encode signals
drawn from the distribution $P[x(t)]$ if we use coding strategies that are
optimized for $Q[x(t)]$. 
If $x(t)$ is a written text, two readers who expect different numbers of spelling errors  will
have  different $Q$s, but to the extent that spelling errors can be corrected by reference to the 
immediate neighboring letters
we insist that any measure of complexity be invariant to these differences in 
$Q$.
On the other hand,
readers who differ in their expectations about the global subject of the 
text may well
disagree about the richness of a newspaper 
article.
This suggests that complexity is a component of the relative 
entropy
that is invariant under some class of `local translations and 
misspellings.'

Suppose that we leave aside global expectations, and construct our 
reference distribution $Q[x(t)]$ by
allowing only for  short ranged interactions---certain letters tend 
to follow
one another, letters form words, and so on, but we bound the range over 
which these rules
are applied.  Models of this
class cannot embody the full structure of most interesting time series (including language), but
in the present  context we are not
asking for this.  On the contrary, we are looking for a measure that is
invariant to differences in this short ranged structure.  In the 
terminology of
field theory or statistical mechanics, we are constructing  our reference 
distribution
$Q[x(t)]$ from local operators.  Because we are 
considering a one dimensional
signal (the one dimension being time), distributions constructed from local 
operators
cannot have any phase transitions as a function of parameters, and the 
absence of
critical points means that the entropy of these distributions (or their 
contribution to
the relative entropy) consists of an extensive term (proportional to the time 
window $T$) plus a
constant subextensive term, plus terms that vanish as $T$ becomes large. 
 Thus,
if we choose different reference distributions within the class 
constructible from
local operators, we can change the extensive component of the relative 
entropy, and
we can change constant subextensive terms, but the divergent subextensive 
terms
are invariant.

To summarize, the usual constraints on information measures in the 
continuum produce
a family of allowable measures, the relative entropy to an arbitrary reference 
distribution.
If we insist that all observers who choose reference distributions 
constructed from
local operators arrive at the same measure of complexity, then 
this measure
must be the divergent subextensive component of the entropy.   As 
emphasized above, the predictive information is the subextensive 
component of
the entropy.   Thus, if we accept the invariance principle as a supplement 
to Shannon's
original postulates, the unique measure of complexity or richness of a signal is the 
divergent component
of the predictive information.  We have seen that this component is  
connected
to learning, quantifying the amount that can be learned about dynamics that generate
the signal, and to
measures of  complexity
that have arisen in statistics and dynamical systems theory.

We thank V. Balasubramanian, A. Bell, S. Bruder, C. Callan, R. Koberle,
A. Libchaber, I. Nemenman, S. Still, and S. Strong  for many helpful discussions.

\end{multicols}
\end{document}